# Amplification and attenuation of light in a waveguide modulated by a travelling wave

M. SUMETSKY

*Aston Institute of Photonic Technologies, Aston University, Birmingham B4 7ET, UK*
*m.sumetsky@aston.ac.uk*

**Abstract:** Light propagating in an optical waveguide can gain or lose power through interaction with a travelling acoustic wave or radio-frequency modulation of permittivity. Here, we model this propagation by considering an optical wave interacting with a weak travelling-wave permittivity perturbation whose frequency is much smaller than the optical frequency and whose amplitude decays exponentially along the propagation direction. Four modulation cases are analyzed: instantaneous modulation, synchronous modulation, and the Stokes and anti-Stokes resonances. For these cases, simple expressions are obtained for the carrier and sideband transmission and reflection powers as well as for the total gain and loss powers. Although the achievable total gain remains small for realistic modulation and waveguide parameters, the anti-Stokes resonance is identified as the most promising condition for observing modulation-induced light amplification.

## 1. Introduction

Brillouin scattering [1, 2] and spatiotemporal permittivity modulation [3-5] represent two ways to the generation of sidebands and amplification of light, both based on the energy exchange between the optical waves and nonstationary permittivity perturbation. In Brillouin amplification, a strong optical pump excites a coherent acoustic wave through electrostriction. This acoustic wave forms a travelling wave modulation of the permittivity that enables narrowband amplification or attenuation of light, as well as the generation of frequency-shifted sidebands [1, 2].

In contrast, when the time-dependent permittivity perturbation is induced by external sources such as an interdigital transducer (IDT) [6, 7] and an RF capacitor [8, 9], the modulation can be maintained independently of the optical fields. This modulation directly couples optical modes separated in frequency and wavenumber, enabling amplification and attenuation, as well as frequency sideband and optical frequency-comb generation [3-5].

A systematic theory of time-modulated electromagnetic fields was first developed in the mid-20[th] century in the context of microwave and guided-wave electronics, including purely time modulation and travelling wave modulation [10-23]. These works established the fundamentals of the travelling-wave modulation theory based on coupled Floquet harmonics, dispersion equation, and phase matching conditions. The simplest wave equation commonly considered in these studies has the form:

$$\frac{\partial^2 (\varepsilon E)}{\partial t^2} - c^2 \frac{\partial^2 E}{\partial x^2} = 0 . \qquad (1)$$

Here $E(x,t)$ is the electromagnetic field, $c$ is the speed of light, and the dependence of permittivity $\varepsilon(x,t)$ on time $t$ and coordinate $x$ has the form of a travelling wave with frequency $\omega_p$ and propagation constant $k_p$,

$$\varepsilon(x,t) = \varepsilon_0 + \Delta\varepsilon(x,t), \quad \Delta\varepsilon(x,t) = \Delta\varepsilon_p \cos\left(k_p x - \omega_p t\right). \qquad (2)$$

In Ref. [16] solution of Eq. (1) was analyzed by perturbation over the modulation amplitude $\Delta\varepsilon_p \ll \varepsilon_0$, while the modulation part of permittivity, $\Delta\varepsilon(x,t)$ was defined by Eq. (2) in the region $0 < x < L$ and assumed to be zero outside it. The coherent wave with frequency $\omega_0$ and propagation constant $k_0$ incident from the left-hand side at $x < 0$ onto the modulated region was set to

$$E^{(0)}(x,t) = \exp(ik_0 x - i\omega_0 t). \tag{3}$$

Three peculiar cases of the resonant transmission and reflection were considered. The first two are the *Stokes* and *anti-Stokes resonances* corresponding, respectively, to sign + and – in the equation

$$k_p = k_p^\pm = \left(\pm 2 - \frac{\omega_p}{\omega_0}\right) k_0. \tag{4}$$

For the stationary case, $\omega_p = 0$, Eq. (2) describes a regular grating, and Eq. (4) corresponds to the Bragg resonance. The third case considered in Ref. [16] is the *synchronous resonance*, when the phase velocity of modulation, $v_p$, is equal to the phase velocity of the incident wave, $v_0$:

$$v_p = v_0, \quad v_0 = \frac{\omega_0}{k_0}, \quad v_p = \frac{\omega_p}{k_p}. \tag{5}$$

The authors of Ref. [17] showed that the power series expansion in $\Delta\varepsilon_p$ diverges in the vicinity of the synchronous resonance and, thus, the perturbation theory proposed in Ref. [16] cannot be applied. Alternatively, the authors of Refs. [14, 15, 23] solved the problem in the WKB (eikonal) approximation. This approximation assumes that the modulating wave has a frequency and propagation constant much smaller than those of the input wave, $\omega_p \ll \omega_0$ and $k_p \ll k_0$. The approach developed in Refs. [14, 15, 23] is not based on the perturbation over $\Delta\varepsilon_p$ and, therefore, is valid in the synchronous case defined by Eq. (5). As a result, it was shown that significant amplification of the input wave power interacting with the travelling wave in the synchronous regime is possible. It was also shown that, in the case of a system having permeability and permittivity proportionally varied in time and space, i.e., possessing a constant impedance, the electromagnetic wave equations can be solved exactly, and only the forward propagating wave can be generated by the input wave of Eq. (3) [10, 14].

Recently, the spatiotemporal concepts described have been revisited and further developed in both microwave theory [3, 5] and photonics [4]. The transmission nonreciprocity, optical frequency comb generation, and amplification effects in space-time-modulated waveguides were detailly investigated [24-32]. Conservation laws for the wave equation with permittivity defined by Eq. (2) and flux invariants governing redistribution among space-time harmonics were formulated and analyzed [33-35].

Of special interest became the regime of slow travelling-wave modulation, $\omega_p \ll \omega_0$ [24-32]. Earlier results on the transformation and amplification of electromagnetic waves by low-frequency modulation [10-23] have been advanced and applied to the propagation of light [29-32]. However, the analysis was primarily applied to the systems with parameters satisfying the eikonal conditions, $\omega_p \ll \omega_0$ and $k_p \ll k_0$ [32] or to systems with constant impedance [30], while the important cases of Stokes and anti-Stokes resonance propagation when $k_p \sim k_0$ (see Eq. (4)) have not been fully studied.

To the author's knowledge, the solution of the wave equation, Eq. (1), has been investigated so far primarily for the travelling wave modulation that is spatially uniform in a finite or infinite region, where it is defined by Eq. (2). For this model, solving Eq. (1) by perturbations in $\Delta\varepsilon_p/\varepsilon_0$ leads to solution divergence near the synchronous, Stokes, and anti-Stokes resonances [17, 19]. However, in practice, surface and bulk acoustic travelling waves are not uniform. For example, an acoustic wave launched by an IDT along an optical waveguide decays exponentially with the waveguide length, and achieving spatial uniformity is challenging. Consequently, it is interesting to investigate the effect of modulation nonuniformity on light propagation.

Here, we investigate the propagation of light in an optical waveguide modulated by a travelling wave *attenuating in space*. Besides its practical importance resembling the behavior of an IDT-generated acoustic wave, the introduction of travelling-wave attenuation allows us to avoid the divergence problem [17, 19, 28] and to derive simple, physically transparent expressions for the transmission and reflection resonances of the carrier and sidebands. Below, these expressions are obtained within the second-order perturbation theory in $\Delta\varepsilon_p/\varepsilon_0$ for instantaneous, synchronous, Stokes, and anti-Stokes modulation, as well as for the corresponding *total power gain and loss*. In particular, we present realistic estimates for the possible total amplification effect for light interacting with an acoustic wave. Generally, this effect is small. However, for a substantial modulation length, it can be distinctly larger at the anti-Stokes resonance than at the Stokes resonance, synchronous modulation, or instantaneous modulation.

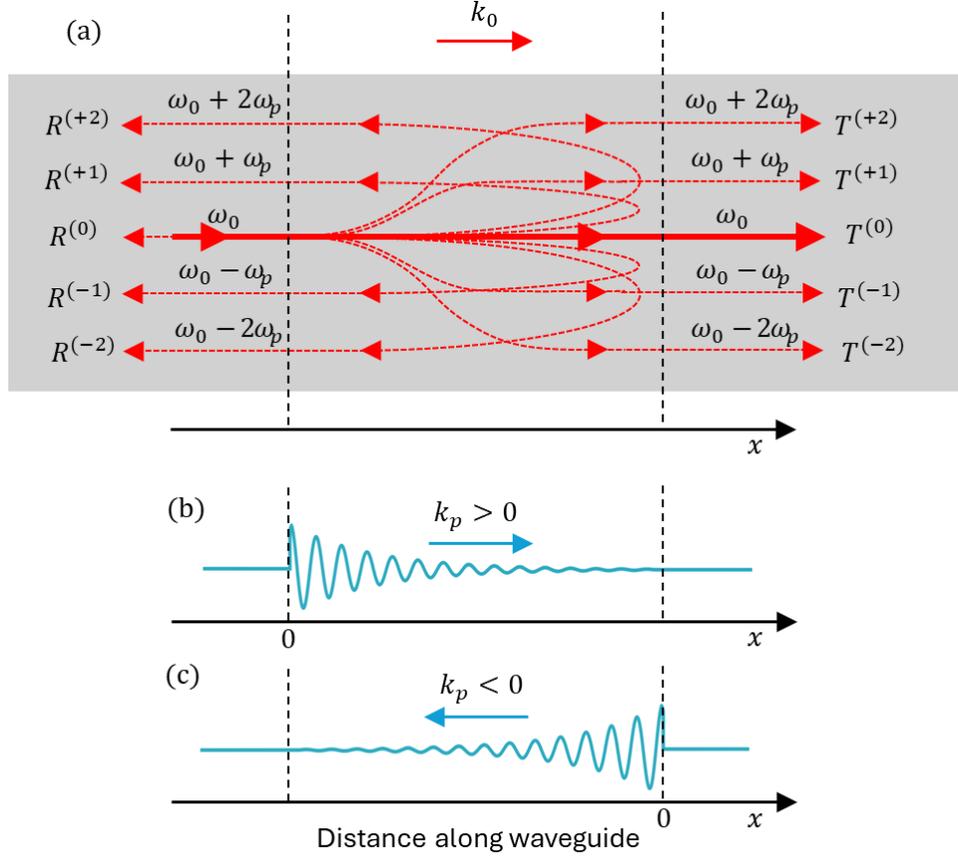

**Fig. 1.** Illustration of light propagating along an optical waveguide modulated by a spatially attenuating travelling wave. (a) A waveguide supporting the carrier optical wave with frequency $\omega_0$ and sideband waves with frequencies $\omega_0 \pm \omega_p$ and $\omega_0 \pm 2\omega_p$. (b) A traveling wave co-propagating with the input optical wave ($k_p/k_0 > 0$). A travelling wave counter-propagating with respect to the input optical wave ($k_p/k_0 < 0$).

## 2. General formalism

Propagation of light along the photonic waveguide is described by the wave equation, Eq. (1). In contrast to the spatially uniform travelling wave modulation of permittivity defined by Eq. (2), we set its dependence on time $t$ and coordinate $x$ along the waveguide as

$$\varepsilon(x,t) = \varepsilon_0 + \Delta\varepsilon(x,t), \quad \Delta\varepsilon(x,t) = \Delta\varepsilon_p \exp(-\alpha_p x)\cos(k_p x - \omega_p t) H(x), \tag{6}$$

where $H(x)$ is the Heaviside function and $\alpha_p$ determines the spatial attenuation of the travelling wave. We assume that a coherent optical wave defined by Eq. (3) is incident from the left-hand side ($x < 0$) into the modulation region. The waveguide considered below, which supports the carrier optical wave with frequency $\omega_0$ and sideband waves with frequencies $\omega_0 \pm \omega_p$ and $\omega_0 \pm 2\omega_p$, is illustrated in Fig. 1(a). The traveling waves co-propagating ($k_p/k_0 > 0$) and counter-propagating ($k_p/k_0 < 0$) with respect to the input optical wave are illustrated in Figs. 1(b) and 1(c). The spatial profile of attenuation in Eq. (6) resembles the attenuation of surface and bulk acoustic waves, in particular, those generated by IDTs [7, 8]. As shown below, the presence of exponential attenuation allows us to directly apply the perturbation theory to Eq. (1) with the space-time dependent permittivity defined by Eq. (6) and to avoid the singularity and divergence problems at the synchronous resonance as well as at the Stokes and anti-Stokes resonances [17, 19, 28]. In particular, rather than expanding the solution into a series of regular space-time harmonics [17, 19, 28], we expand it into harmonics attenuating in space.

Since we are interested in arbitrary relations between the propagation constants $k_0$ and $k_p$, and, in particular, the case $k_0 \sim k_p$, the eikonal and conventional coupled wave equation theories cannot be applied to solve the wave equation, Eq. (1). Instead, we solve this equation by perturbation over $\Delta\varepsilon_p/\varepsilon_0$:

$$E(x,t)\big|_{x>0} = E^{(0)}(x,t) + E^{(1)}(x,t) + E^{(2)}(x,t) + \ldots + \sum_{\pm} \Phi^{(\pm)}(\omega_0 t \pm k_0 x) \ . \tag{7}$$

Here, $\Phi^{(\pm)}(\omega_0 t)$ are arbitrary functions with small magnitude, $|\Phi^{(\pm)}(\omega_0 t)| \ll 1$, and functions $E^{(m)}(x,t)$ are found from the recurrent equations:

$$\varepsilon_0 \frac{\partial^2 E^{(m)}}{\partial t^2} - c^2 \frac{\partial^2 E^{(m)}}{\partial x^2} = -\frac{\partial^2 \left(\Delta\varepsilon E^{(m-1)}\right)}{\partial t^2} \ , \tag{8}$$

where the zero-order solution, $E^{(0)}(x,t)$, coincides with the incident wave defined by Eq. (3). While $E^{(0)}(x,t) \sim 1$ and depends on time as $\exp(-i\omega_0 t)$, the term $E^{(m)}(x,t)$ in Eq. (7) is proportional to $(\Delta\varepsilon_p/\varepsilon_0)^m$, decays with growing $x$, and includes harmonics depending on time as $\exp(-i(\omega_0 + q\omega_p)t)$, $q = -m, -m+1, \ldots, m$:

$$E^{(m)}(x,t) = \left(\frac{\Delta\varepsilon_p}{\varepsilon_0}\right)^m \exp(ik_0 x - i\omega_0 t - m\cdot\alpha x) \sum_{q=-m}^{m} \left(U_q^{(m)} \exp(iq\cdot(k_p x - \omega_p t))\right). \tag{9}$$

with constant parameters $U_q^{(m)}$. Substitution of Eq. (9) into Eq. (8) reduces the latter to an ordinary differential equation, which is solved analytically and allows to determine $U_q^{(m)}$.

Due to the absence of ingoing waves on the right-hand side at $x \gg \alpha_p^{-1}$, function $\Phi^{(-)}(t)$ in Eq. (7) is set to zero at $x > 0$, and the function $\Phi^{(+)}(t)$ in this region is assumed to be a sum of outgoing waves:

$$\Phi^{(+)}(x,t) = \sum_m T^{(m)} \exp\left(i\left(1 + m\cdot\frac{\omega_p}{\omega_0}\right)(k_0 x - \omega_0 t)\right). \tag{9}$$

Here, the partial (sideband) *transmission coefficients* $T^{(m)}$ are proportional to $(\Delta\varepsilon_p/\varepsilon_0)^{|m|}$ and are determined from the condition of continuity of the solution and its derivative at $x=0$ as a series $T^{(m)} = \sum_{q=|m|}^{\infty}((\Delta\varepsilon_p/\varepsilon_0)^{|m|}T_q^{(m)})$. Similarly, at $x < 0$, we set $\Phi^{(+)}(t) = 0$ and

$$\Phi^{(-)}(x,t) = \sum_m R^{(m)} \exp\left(i\left(1 + m\cdot\frac{\omega_p}{\omega_0}\right)(-k_0 x - \omega_0 t)\right), \tag{10}$$

where the sideband *reflection coefficients* $R^{(m)}$ are proportional to $(\Delta\varepsilon_p/\varepsilon_0)^m$ and can be determined from the same continuity condition at $x = 0$.

As a result of the interaction of the optical wave with the traveling wave, the optical wave can gain or lose power. Therefore, the *total output power gain or loss*, which is determined as

$$\Delta P^{(tot)} = \sum_m \left(\left|T^{(m)}\right|^2 + \left|R^{(m)}\right|^2\right) - 1 , \tag{11}$$

can be positive, indicating power gain, or negative, indicating power loss. Below, we use the above equations to determine $\Delta P^{(tot)}$ as well as the carrier and sideband transmission and reflection powers, $|T^{(0)}|^2$, $|T^{(\pm 1)}|^2$, $|R^{(0)}|^2$, and $|R^{(\pm 1)}|^2$ to *second order in* $\Delta \varepsilon_p / \varepsilon_0$. Since $T^{(\pm 1)}$, $R^{(0)}$, and $R^{(\pm 1)}$ are of the first order in $\Delta \varepsilon_p / \varepsilon_0$, it is sufficient to calculate them to the first order to determine the corresponding powers to the second order. In contrast, $T^{(0)}$ should be calculated to second order in $\Delta \varepsilon_p / \varepsilon_0$.

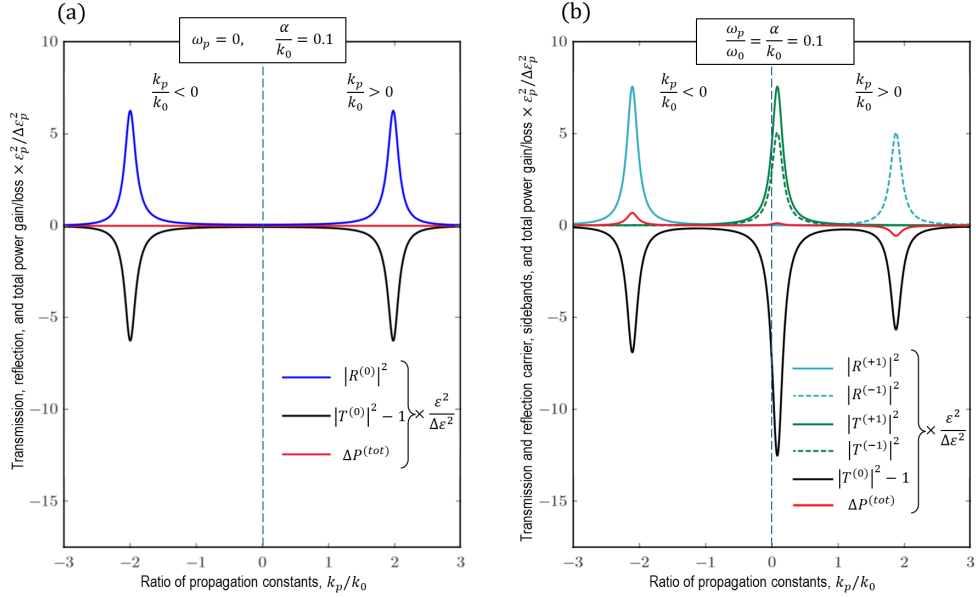

**Fig. 2.** (a) Normalized transmission and reflection power for a stationary grating as a function of $k_p/k_0$ for $\alpha_p/k_0 = 0.1$. (b) Normalized transmission and reflection power for the carrier and sidebands, and total power gain as a function of $k_p/k_0$ for modulation by a traveling wave with relative parameters $\omega_p/\omega_0 = \alpha_p/k_0 = 0.1$.

## 3. Stationary modulation

For comparison, we first consider the stationary modulation (i.e., regular spatial grating) when $\omega_p = 0$. Then the transmission and reflection coefficients $T_{stat}^{(m)} = R_{stat}^{(m)} = 0$ for $|m| > 0$, and $P_{stat}^{(tot)} = |T_{stat}^{(0)}|^2 + |R_{stat}^{(0)}|^2 = 1$. In particular, to the second-order perturbation over $\Delta \varepsilon_p / \varepsilon_0$, we have:

$$\left|R_{stat}^{(0)}\right|^2 = 1 - \left|T_{stat}^{(0)}\right|^2 = \left(\frac{\Delta \varepsilon_p}{2\varepsilon_0}\right)^2 \frac{k_0^2 \left(4k_0^2 + \alpha_p^2\right)}{\left(\left(2k_0 - k_p\right)^2 + \alpha_p^2\right)\left(\left(2k_0 + k_p\right)^2 + \alpha_p^2\right)}, \qquad (12)$$

From this equation, the normalized reflection power $|R_{stat}^{(0)} \Delta \varepsilon_p / \varepsilon_0|^2$ can be completely expressed through dimensionless parameters $\alpha_p/k_0$ and $k_p/k_0$. It achieves maximum at the well-known Bragg resonance when [36]

$$k_p = k_p^\pm = \pm 2k_0. \qquad (13)$$

As a characteristic example, Fig. 2(a) shows the plot of $|R_{stat}^{(0)} \Delta \varepsilon_p / \varepsilon_0|^2$ as a function of $k_p/k_0$ for $\alpha_p/k_0 = 0.1$. Table 1 presents the expressions for the transmission and reflection powers at the Bragg resonance and also in the absence of the grating, for $k_p = 0$. In the latter case, the appearance of reflection and deviation of transmission from unity is caused by the discontinuity of permittivity at $x = 0$.

**Table 1. Transmission and reflection power of a stationary grating ($\omega_p = 0$) at Bragg resonance ($k_p = \pm k_0$) and in the absence of grating ($k_p = 0$).**

| Reflection and transmission gain/loss | Backward Bragg resonance, $k_p = \pm k_0$ | No grating, $k_p = 0$ |
|---|---|---|
| $\|R\|^2$ | $\frac{1}{16}\left(\frac{k_0}{\alpha_p}\right)^2\left(\frac{\Delta\varepsilon_p}{\varepsilon_0}\right)^2$ | $\frac{1}{16}\left(\frac{\Delta\varepsilon_p}{\varepsilon_0}\right)^2$ |
| $\|T\|^2 - 1$ | $-\frac{1}{16}\left(\frac{k_0}{\alpha_p}\right)^2\left(\frac{\Delta\varepsilon_p}{\varepsilon_0}\right)^2$ | $-\frac{1}{16}\left(\frac{\Delta\varepsilon_p}{\varepsilon_0}\right)^2$ |
| Total power gain/loss $\Delta P^{(tot)}$ | 0 | 0 |

## 4. Travelling wave modulation

At non-zero modulation frequency, $\omega_p \neq 0$, the transmission and reflection amplitudes of the carrier and first-order sidebands, calculated to the first order in $\Delta\varepsilon_p/\varepsilon_0$, are:

$$R^{(0)} = 0, \tag{14}$$

$$T^{(\pm 1)} = \frac{\Delta\varepsilon_p}{2\varepsilon_0} \cdot \frac{k_0(\omega_0 \pm \omega_p)}{\pm(k_p\omega_0 - k_0\omega_p) + i\alpha_p\omega_0}, \tag{15}$$

$$R^{(\pm 1)} = \frac{\Delta\varepsilon_p}{2\varepsilon_0} \cdot \frac{k_0(\omega_0 \pm \omega_p)}{2k_0\omega_0 \pm k_p\omega_0 \pm k_0\omega_p + i\alpha_p\omega_0}, \tag{16}$$

which immediately allows us to calculate the transmission and reflection powers $|T^{(\pm 1)}|^2$ and $|R^{(\pm 1)}|^2$. More cumbersome expressions for the carrier transmission power, $|T^{(0)}|^2$, and total power gain/loss $\Delta P^{(tot)}$ were found using the symbolic Mathcad tool. For briefness, we introduce the dimensionless parameters

$$\bar{\omega} = \frac{\omega_p}{\omega_0}, \quad \bar{k} = \frac{k_p}{k_0}, \quad \bar{\alpha} = \frac{\alpha_p}{k_0}. \tag{17}$$

Then we find:

$$|T^{(0)}|^2 = 1 + \left(\frac{\Delta\varepsilon_p}{\varepsilon_0}\right)^2 \frac{A(\bar{\omega},\bar{k},\bar{\alpha})}{8D(\bar{\omega},\bar{k},\bar{\alpha})},$$

$$\Delta P^{(tot)} = \left(\frac{\Delta\varepsilon_p}{\varepsilon_0}\right)^2 \frac{B(\bar{\omega},\bar{k},\bar{\alpha})}{4D(\bar{\omega},\bar{k},\bar{\alpha})}, \tag{18}$$

where

$$A(\bar{\omega},\bar{k},\bar{\alpha}) = -\left(\bar{\alpha}^4 + 6\bar{\alpha}^2 + 8 + (2\bar{\alpha}^2 - 2)\bar{k}^2 + \bar{k}^4\right) + 12\bar{k}\bar{\omega} + (2 - 4\bar{k}^2)\bar{\omega}^2 - 4\bar{k}\bar{\omega}^3 + \bar{\omega}^4$$

$$B(\bar{\omega},\bar{k},\bar{\alpha}) = \bar{\omega}\left[-2\bar{k}(\bar{\alpha}^2 + \bar{k}^2) + ((\bar{\alpha}^2 + \bar{k}^2)^2 + 4\bar{\alpha}^2 + 8)\bar{\omega} + \right.$$
$$\left. 2\bar{k}(\bar{\alpha}^2 + \bar{k}^2 - 5)\bar{\omega}^2 + 2(\bar{\alpha}^2 + \bar{k}^2 - 2)\bar{\omega}^3 + 2\bar{k}\bar{\omega}^4 + \bar{\omega}^5\right] \tag{19}$$

$$D(\bar{\omega},\bar{k},\bar{\alpha}) = \left(\bar{\alpha}^2 + \bar{k}^2 - 2\bar{k}\bar{\omega} + \bar{\omega}^2\right)\left((2-\bar{k})^2 + \bar{\alpha}^2 + (2\bar{k} - 4)\bar{\omega} + \bar{\omega}^2\right) \times$$
$$\left((2+\bar{k})^2 + \bar{\alpha}^2 + (2\bar{k} - 4)\bar{\omega} + \bar{\omega}^2\right)$$

We note that while the case of slow modulation frequency $\bar{\omega} = \omega_p/\omega_0 \ll 1$ is of our major interest here, Eqs. (15)-(19) are valid for any relation between $\omega_p$ and $\omega_0$.

Demonstrative dependencies of the carrier, sideband, and total gain/loss powers as a function of $k_p/k_0$ are presented in Fig. 2(b). For better visualization, in this figure, we set the dimensionless parameters $\omega_p/\omega_0 = \alpha_p/k_0 = 0.1$, which are four orders of magnitude larger than those for realistic acoustic modulation of photonic waveguides to be considered later.

Table 2. Expressions for the carrier and sideband transmission and reflection power at Stokes, anti-Stokes, and synchronous resonances, for instantaneous modulation, and for the total power gain/loss calculated to the second order in $\Delta\varepsilon_p/\varepsilon_0$.

| Sidebands and carrier power gain/loss | Backward Stokes resonance, $k_p = \left(2 - \frac{\omega_p}{\omega_0}\right)k_0$ | Backward Anti-Stokes resonance, $k_p = -\left(2 + \frac{\omega_p}{\omega_0}\right)k_0$ | Forward synchronous resonance, $v_p = v_0$ | Forward instantaneous modulation, $k_p = 0$ |
|---|---|---|---|---|
| $\lvert T^{(+1)}\rvert^2$ | $\frac{1}{64}\left(\frac{\Delta\varepsilon_p}{\varepsilon_0}\right)^2$ | $\frac{1}{64}\left(\frac{\Delta\varepsilon_p}{\varepsilon_0}\right)^2$ | $\frac{1}{16}\left(\frac{k_0}{\alpha_p}\right)^2\left(\frac{\Delta\varepsilon_p}{\varepsilon_0}\right)^2$ | $\frac{1}{16}\frac{k_0^2\omega_0^2}{k_0^2\omega_p^2+\alpha_p^2\omega_0^2}\left(\frac{\Delta\varepsilon_p}{\varepsilon_0}\right)^2$ |
| $\lvert T^{(-1)}\rvert^2$ | $\frac{1}{64}\left(\frac{\Delta\varepsilon_p}{\varepsilon_0}\right)^2$ | $\frac{1}{64}\left(\frac{\Delta\varepsilon_p}{\varepsilon_0}\right)^2$ | $\frac{1}{16}\left(\frac{k_0}{\alpha_p}\right)^2\left(\frac{\Delta\varepsilon_p}{\varepsilon_0}\right)^2$ | $\frac{1}{16}\frac{k_0^2\omega_0^2}{k_0^2\omega_p^2+\alpha_p^2\omega_0^2}\left(\frac{\Delta\varepsilon_p}{\varepsilon_0}\right)^2$ |
| $\lvert R^{(+1)}\rvert^2$ | $\frac{1}{256}\left(\frac{\Delta\varepsilon_p}{\varepsilon_0}\right)^2$ | $\frac{1}{16}\left(\frac{k_0}{\alpha_p}\right)^2\left(\frac{\Delta\varepsilon_p}{\varepsilon_0}\right)^2$ | $\frac{1}{64}\left(\frac{\Delta\varepsilon_p}{\varepsilon_0}\right)^2$ | $\frac{1}{64}\left(\frac{\Delta\varepsilon_p}{\varepsilon_0}\right)^2$ |
| $\lvert R^{(-1)}\rvert^2$ | $\frac{1}{16}\left(\frac{k_0}{\alpha_p}\right)^2\left(\frac{\Delta\varepsilon_p}{\varepsilon_0}\right)^2$ | $\frac{1}{256}\left(\frac{\Delta\varepsilon_p}{\varepsilon_0}\right)^2$ | $\frac{1}{64}\left(\frac{\Delta\varepsilon_p}{\varepsilon_0}\right)^2$ | $\frac{1}{64}\left(\frac{\Delta\varepsilon_p}{\varepsilon_0}\right)^2$ |
| $\lvert T^{(0)}\rvert^2 - 1$ | $-\frac{1}{16}\left(\frac{k_0}{\alpha_p}\right)^2\left(\frac{\Delta\varepsilon_p}{\varepsilon_0}\right)^2$ | $-\frac{1}{16}\left(\frac{k_0}{\alpha_p}\right)^2\left(\frac{\Delta\varepsilon_p}{\varepsilon_0}\right)^2$ | $-\frac{1}{8}\left(\frac{k_0}{\alpha_p}\right)^2\left(\frac{\Delta\varepsilon_p}{\varepsilon_0}\right)^2$ | $-\frac{1}{8}\frac{k_0^2\omega_0^2}{k_0^2\omega_p^2+\alpha_p^2\omega_0^2}\left(\frac{\Delta\varepsilon_p}{\varepsilon_0}\right)^2$ |
| $\lvert R^{(0)}\rvert^2$ | 0 | 0 | 0 | 0 |
| Total power gain/loss $\Delta P^{(tot)}$ | $-\frac{1}{16}\frac{\omega_p}{\omega_0}\left(\frac{k_0}{\alpha_p}\right)^2\left(\frac{\Delta\varepsilon_p}{\varepsilon_0}\right)^2$ | $\frac{1}{16}\frac{\omega_p}{\omega_0}\left(\frac{k_0}{\alpha_p}\right)^2\left(\frac{\Delta\varepsilon_p}{\varepsilon_0}\right)^2$ | $\frac{1}{8}\left(\frac{\omega_p}{\omega_0}\right)\left(\frac{k_0}{\alpha_p}\right)^2\left(\frac{\Delta\varepsilon_p}{\varepsilon_0}\right)^2$ | $\frac{1}{8}\frac{k_0^2\omega_p^2}{k_0^2\omega_p^2+\alpha_p^2\omega_0^2}\left(\frac{\Delta\varepsilon_p}{\varepsilon_0}\right)^2$ |

The behavior of the carrier and sideband transmission and reflection in several particular cases is of special interest. The reflection sideband power $\lvert R^{(-1)}\rvert^2$ exhibits the *Stokes resonance* and the reflection sideband power $\lvert R^{(+1)}\rvert^2$ exhibits the *anti-Stokes resonance* if, respectively, $k_p = k_p^\pm$, where $k_p^\pm$ is defined by Eq. (4). It is seen from Fig. 2(b) that $\lvert R^{(+1)}\rvert^2$ and $\lvert R^{(-1)}\rvert^2$ (solid and dashed blue curves) exhibit positive resonances at, respectively, $k_p = k_p^+$ and $k_p = k_p^-$. At the same positions, the carrier power $\lvert T^{(0)}\rvert^2$ has negative resonances. As a result, the *total power* exhibits a small gain at the anti-Stokes resonance and a small loss at the Stokes resonance. Simple expressions for the carrier, sideband, and total power gain/loss at the Stokes and anti-Stokes resonances following from Eqs. (4), (5), and (14)-(19) are collected in Table 2. It shows that both gain and loss have the same absolute value equal to $(\omega_p/2\omega_0)(k_0^2/\alpha_p^2)(\Delta\varepsilon_p^2/16\varepsilon_0^2)$, i.e., linearly vanish with modulation frequency $\omega_p$ and grow fast with the characteristic modulation length $\alpha_p^{-1}$. Consequently, the most promising condition for the total amplification of light corresponds to the anti-Stokes resonance when (see Table 2)

$$\Delta P^{(tot)} = \frac{1}{16}\frac{\omega_p}{\omega_0}\left(\frac{k_0}{\alpha_p}\right)^2\left(\frac{\Delta\varepsilon_p}{\varepsilon_0}\right)^2. \qquad (20)$$

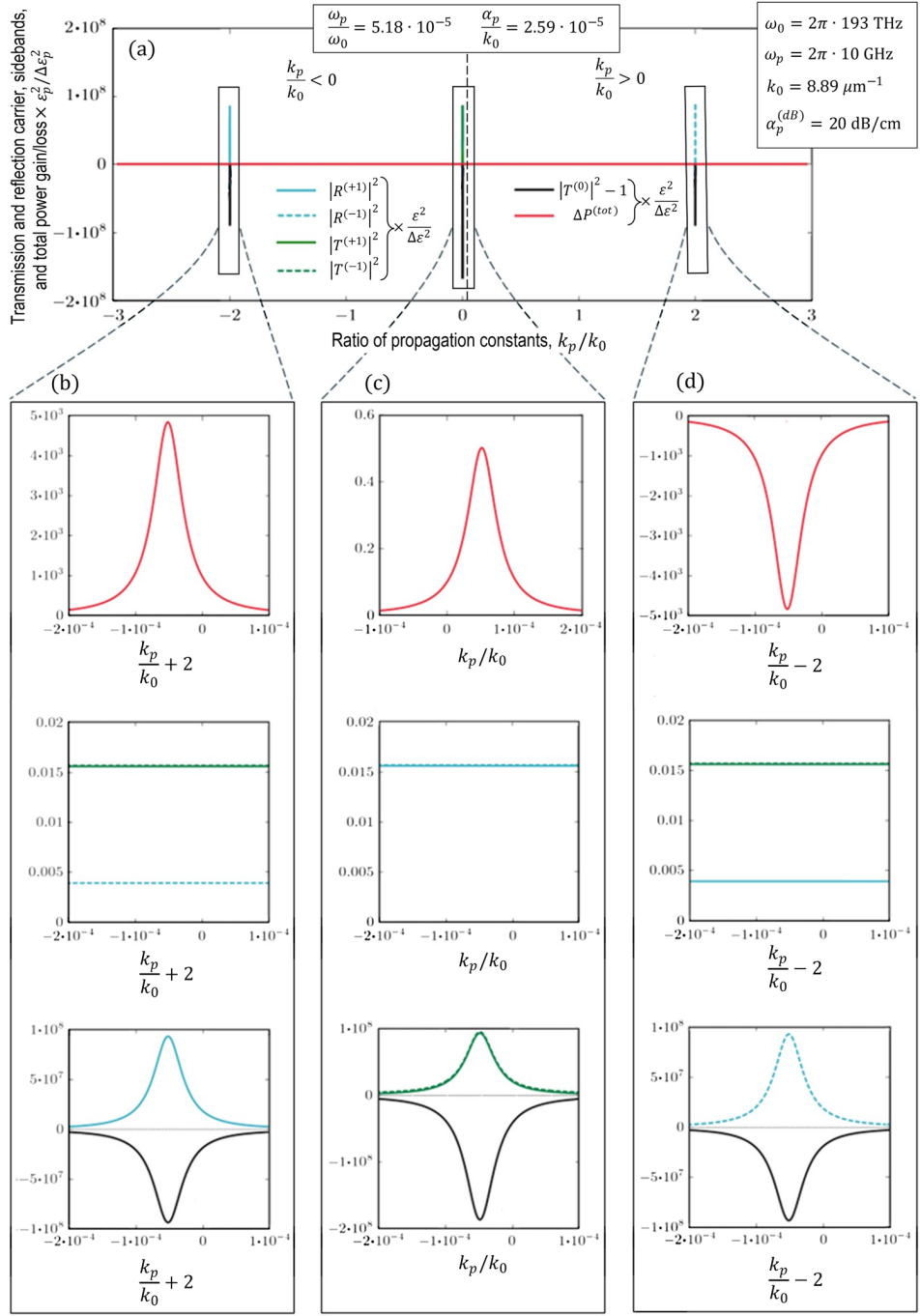

**Fig. 3.** (a) Normalized transmission and reflection power for the carrier and sidebands, and the total power gain/loss as a function of $k_p/k_0$ for optical wave and modulation parameters, $\omega_0 = 2\pi \cdot 193$ THz, $\omega_p = 2\pi \cdot 10$ GHz, $\varepsilon_0 = 4.84$ (lithium niobate), $k_0 = \omega_0 \varepsilon_0^{1/2}/c = 8.89\ \mu m^{-1}$, and $\alpha_p = 230\ m^{-1}$ corresponding, in dBs, to the modulation power attenuation $\alpha_p^{(dB)} = 20$ dB/cm. (b), (c), and (d) Sections of plots shown in (a) expanded near the anti-Stokes, synchronous, and Stokes resonances.

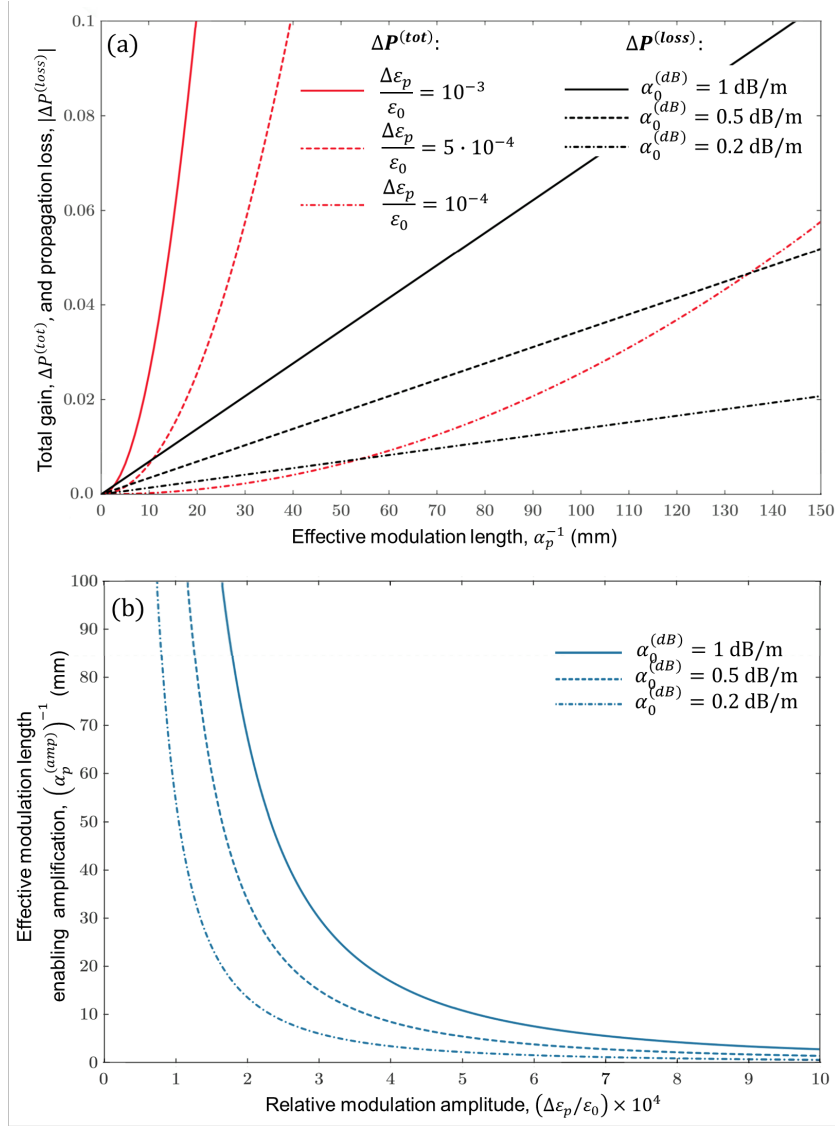

**Fig. 4.** (a) Comparison of modulation-induced total gain, $\Delta P^{(tot)}$, at the anti-Stokes resonance (red solid, dashed, and dash-dotted lines) with material loss-induced attenuation $\Delta P^{(loss)}$ calculated for the waveguide length $3\alpha_p^{-1}$ as a function of effective modulation length $\alpha_p^{-1}$ for different relative permittivity modulation amplitudes $\Delta\varepsilon_p/\varepsilon_0$ and material attenuations $\alpha_0^{(dB)}$. (b) Effective modulation length enabling amplification of light as a function of relative modulation amplitudes $\Delta\varepsilon_p/\varepsilon_0$ for different material attenuations $\alpha_0^{(dB)}$.

Both transmission sideband powers $|T^{(\pm 1)}|^2$ (solid and dashed green curves in Fig. 2(b)) exhibit the *synchronous resonance* (Eq. (5) and Table 2) if the optical wave phase velocity, $v_0$, coincides with the traveling wave phase velocity, $v_p$. Similar to Stokes and anti-Stokes resonances, the carrier power $|T^{(0)}|^2$ at the condition of Eq. (4) has a negative resonance, while the resonance value for both $|T^{(+1)}|^2$ and $|T^{(-1)}|^2$ is positive and equal to $(\omega_p^2/\omega_0^2)(k_0^2/\alpha_p^2)(\Delta\varepsilon_p^2/8\varepsilon_0^2)$. Consequently, these resonances vanish much faster with modulation frequency

$\omega_p$ (as $\omega_p^2$ rather than as $\omega_p$), compared to the Stokes and anti-Stokes resonances, and grow similarly, proportionally to the characteristic modulation length $\alpha_p^{-1}$ squared.

In the *stationary* limit, $\omega_p \to 0$, we find $R^{(+1)} + R^{(-1)} = R_{stat}^{(0)}$. Then, the sideband reflection power $|R^{(\pm 1)}|^2$ (see Eq. (16) and Table 2) at the anti-Stokes and Stokes resonances coincides with the stationary reflection power $|R_{stat}^{(0)}|^2$ (Eq. (12) and Table 1) at the Bragg resonance. The corresponding expressions for the carrier, sideband, and total power gain/loss are given in Table 2. It is seen from this Table that the total power gain vanishes with the modulation frequency as $\omega_p^2$ and, for a large modulation length $\alpha_p^{-1}$, achieves the maximum limit of $\Delta\varepsilon_p^2/8\varepsilon_0^2$.

Fig. 3 shows the dependencies of carrier, sideband, and total gain/loss power on the propagation constant ratio $k_p/k_0$ for the realistic optical wave and modulation parameters, $\omega_0 = 2\pi \cdot 193$ THz, $\omega_p = 2\pi \cdot 10$ GHz, $\varepsilon_0 = 4.84$ (lithium niobate), $k_0 = \omega_0\varepsilon_0^{1/2}/c = 8.89$ µm$^{-1}$, and $\alpha_p = 230$ m$^{-1}$ corresponding, in dBs, to modulation power attenuation $\alpha_p^{(dB)} = 20\log(e)\,\alpha_p = 20$ dB/cm. It is seen that the anti-Stokes, Stokes, and synchronous resonances become dramatically narrower and larger than those in Fig. 2(b). The total power gain at the synchronous resonance remains similar, due to a similar magnitude order of the factor $(\omega_p^2/\omega_0^2)(k_0^2/\alpha_p^2)$ (see Table 2). However, the total normalized power gain at the anti-Stokes resonance, which is proportional to $(\omega_p/\omega_0)(k_0^2/\alpha_p^2)$, becomes more than three orders of magnitude greater ($\sim 5 \cdot 10^3$ in Fig. 3 vs. $\sim 1$ in Fig. 2(b)). Thus, the anti-Stokes resonance condition is the most effective way to achieve light amplification.

To estimate feasible light amplification, we compare the total power gain, $\Delta P^{(tot)}$, at the anti-Stokes resonance with power attenuation, $\Delta P^{(loss)}$, caused by material losses. Since the perturbation theory considered determines only a small gain or loss, $|\Delta P^{(tot)}| \ll 1$, we assume that the material attenuation, $\Delta P^{(loss)}$, is also small and can be calculated for a waveguide with length $L$ and attenuation constant $\alpha_0$ as

$$\Delta P^{(loss)} = -2\alpha_0 L, \quad 2\alpha_0 L \ll 1. \tag{21}$$

Assuming that the modulated waveguide length is $L = 3\alpha_p^{-1}$, we find that the gain $\Delta P^{(tot)}$ will be achieved in this waveguide with a relative precision of $\exp(-2\alpha_p L) = e^{-6} = 0.002$. To overcome losses, the amplification along this waveguide length should satisfy the condition $\Delta P^{(tot)} + \Delta P^{(loss)} > 0$. Using the expression for the $\Delta P^{(tot)}$ at the anti-Stokes resonance of Eq. (20) and the expression for $\Delta P^{(loss)}$ of Eq. (21) with $L = 3\alpha_p^{-1}$, the latter inequality can be written down as

$$\alpha_p < \alpha_p^{(amp)}, \quad \alpha_p^{(amp)} = \frac{k_0^2}{48\alpha_0}\frac{\omega_p}{\omega_0}\left(\frac{\Delta\varepsilon_p}{\varepsilon_0}\right)^2, \tag{22}$$

where the largest modulation attenuation $\alpha_p^{(amp)}$ (and, consequently, the smallest effective modulation length $1/\alpha_p^{(amp)}$) enabling amplification of light is introduced.

As an example, in Fig. 4(a), we compare the dependencies of the power gain $\Delta P^{(tot)}$ at the anti-Stokes resonance and loss $|\Delta P^{(loss)}|$ defined by Eqs. (20) and (21) as a function of effective modulation length $1/\alpha_p$ for the parameters considered above: $\omega_0 = 2\pi \cdot 193$ THz, $\omega_p = 2\pi \cdot 10$ GHz, $\varepsilon_0 = 4.84$, and $k_0 = \omega_0\varepsilon_0^{1/2}/c = 8.89$ µm$^{-1}$. The gain $\Delta P^{(tot)}$ is considered for the amplitudes of relative permittivity modulation $\Delta\varepsilon_p/\varepsilon_0 = 10^{-4}, 5 \cdot 10^{-4}$, and $10^{-3}$. Based on the recently achieved propagation loss of thin film lithium niobate waveguides as small as $\alpha_0^{(dB)} = 20\log(e)\,\alpha_0 = 0.34$ dB/m by chemo-mechanical polishing in Ref. [37], $\alpha_0^{(dB)} = 0.2$ dB/m by post-fabrication annealing in oxygen atmosphere in Ref. [38], and $\alpha_0^{(dB)} = 1.3$ dB/m by robust dry reactive ion etching in Ref. [39], we calculate the waveguide loss $|\Delta P^{(loss)}|$ in Fig. 4(a), for $\alpha_0^{(dB)} = 0.2,\ 0.5,$ and 1 dB/m. It is seen from this figure that, for realistic modulation lengths, the amplitude of relative permittivity modulation $\Delta\varepsilon_p/\varepsilon_0$, enabling overall amplification has to have an order of $10^{-4}$ or larger. It is also seen from Fig. 4(a), that the characteristic waveguide effective length $1/\alpha_p^{(amp)}$ enabling amplification can be reasonably small, $\sim 1$ mm at $\Delta\varepsilon_p/\varepsilon_0 = 10^{-3}$ and rapidly grows with decreasing the modulation amplitude, being $\sim 100$ mm at $\Delta\varepsilon_p/\varepsilon_0 = 10^{-4}$. The actual dependence of $1/\alpha_p^{(amp)}$ on $\Delta\varepsilon_p/\varepsilon_0$ is presented in Fig. 4(b).

## 5. Conclusions

In summary, we analyzed the propagation of an optical wave with frequency $\omega_0$ and propagation constant $k_0$ in a waveguide whose permittivity is perturbed by a travelling wave with a relatively small frequency $\omega_p \ll \omega_0$, propagation constant $k_p$, and attenuation along the waveguide length defined by the attenuation constant $\alpha_p$. Within the second-order perturbation theory in the relative modulation amplitude of the waveguide permittivity, $\Delta\varepsilon_p/\varepsilon_0$, we derived simple and physically transparent expressions for the transmission and reflection powers of the carrier and first-order sidebands, as well as for the total optical power gain and loss.

In particular, we considered the cases of instantaneous modulation ($k_p = 0$), synchronous modulation ($\omega_p/k_p = \omega_0/k_0$), and modulation at the Stokes ($k_p \cong 2k_0$) and anti-Stokes ($k_p \cong -2k_0$) resonances. We showed that, for the instantaneous modulation, the relative power gain is limited by $\Delta\varepsilon_p^2/8\varepsilon_0^2$. In contrast, the gain grows with the modulation length $\alpha_p^{-1}$ and frequency ratio $\omega_p/\omega_0$ and is determined as $\Delta P^{(tot)} = (\omega_p^2/\omega_0^2)(k_0^2/\alpha_p^2)(\Delta\varepsilon_p^2/8\varepsilon_0^2)$ (Table 2) at the synchronous resonance and as $\Delta P^{(tot)} = (\omega_p/\omega_0)(k_0^2/\alpha_p^2)(\Delta\varepsilon_p^2/16\varepsilon_0^2)$ (Eq. (20) and Table 2) at the anti-Stokes resonance. In all cases, the power gain produced by realistic modulation parameters is relatively small. However, for a substantial modulation length, it can become distinctly larger at the anti-Stokes resonance. As a result, the anti-Stokes resonance is suggested as the most favorable configuration for observing travelling-wave-induced optical amplification.

Using realistic parameters for thin-film lithium niobate waveguides modulated by an acoustic wave, we show that modulation-induced optical gain can, in principle, overcome intrinsic waveguide loss when the modulation is sufficiently strong and extended. In particular, for relative permittivity modulation amplitudes in the range $\Delta\varepsilon_p/\varepsilon_0 \sim 10^{-4} - 10^{-3}$, overall amplification requires effective modulation lengths that increase rapidly from the millimeter scale to the ten-centimeter scale as $\Delta\varepsilon_p/\varepsilon_0$ decreases across this range.

The model of spatially attenuating modulation considered here is a particular example of aperiodic spatial modulation. Exploring more general aperiodic spatio-temporal dependencies, and, in particular, those achievable in practice [7, 8, 40], as well as the effect of spatio-temporal noise on light propagation, is an interesting topic for future investigation. We also suggest that extending the present results to optical resonators, where the anti-Stokes interaction can be resonantly enhanced, may enable the realization of devices that exhibit significant light amplification. In contrast to broadband amplification schemes [14, 15, 23, 28–32], resonant anti-Stokes amplification relies on only two harmonic interactions, producing forward-propagating light at the carrier frequency $\omega_0$ and backward-propagating light at the sideband frequency $\omega_0 + \omega_p$.

**Funding.** This research was supported by the Engineering and Physical Sciences Research Council (Grants EP/W002868/1 and EP/X03772X/1).

**Disclosures.** The authors declare no conflicts of interest.

**Data availability.** Data underlying the results presented in this paper are not publicly available at this time but may be obtained from the author upon reasonable request.